\documentclass[onecolumn, prE] {revtex4}
\usepackage{graphics}
\usepackage{epsf}
\usepackage{amsmath}
\usepackage{amssymb}
\usepackage{amscd}
\usepackage{color}
\usepackage{pstricks}
\usepackage{listings} \newcommand{\elabel}[1]{\label{e:#1}}
\newcommand{\slabel}[1]{\label{s:#1}}
\newcommand{\eq}[1]{Eq.~(\ref{e:#1})}
\newcommand{\eqq}[1]{Equation~(\ref{e:#1})}
\newcommand{\eqtwo}[2]{Eqs~(\ref{e:#1}) and~(\ref{e:#2})}
\newcommand{\fig}[1]{Fig.~\ref{f:#1}}
\newcommand{\figg}[1]{Figure~\ref{f:#1}}
 
\newcommand{\sect}[1]{Section~\ref{s:#1}} 
\newcommand{\app}[1]{Appendix~\ref{s:#1}} 
\newcommand{\quot}[1]{``#1''}

\newcommand{\CCAL}{\mathcal{C}}  
\newcommand{\HCAL}{\mathcal{H}}  
\newcommand{\NCAL}{\mathcal{N}}  
\newcommand{\OCAL}{\mathcal{O}}  
\newcommand{\PCAL}{\mathcal{P}}  

\newcommand{\sigmavec}{\boldsymbol{\sigma}}

\newcommand{\expa}[1]{\mathrm{e}^{#1}}   
\newcommand{\expb}[1]{\exp \glb #1 \grb} 
\newcommand{\expc}[1]{\exp \glc #1 \grc} 
\newcommand{\logb}[2][]{\log^{#1} \glb #2 \grb}  

\newcommand{\ran}[2]{\text{ran}[#1,#2]}

\newcommand{\glb}{\left(}  
\newcommand{\grb}{\right)}  
\newcommand{\glc}{\left[}  
\newcommand{\grc}{\right]}  

\newcommand{\TO}{,\ldots,}

\newcommand{\mean}[1]{\left\langle #1 \right\rangle}

\newcommand{\forward}{\text{fw}}  
\newcommand{\backward}{\text{bw}}  
\newcommand{\map}{\text{fw,corr}} 
\newcommand{\mc}{1,1}  
\newcommand{\taucoup}{\tau_{\text{coup}}}  
\newcommand{\tcoup}{t_{\text{coup}}}  
\newcommand{\taucorr}{\tau_{\text{corr}}}  

\newcommand{\Tries}{\ensuremath{\CCAL}}
\newcommand{\Tstart}{T_{\text{start}}}
\newcommand{\wfigure}[2]
   {\begin{figure}[htbp]
   \centerline{
      \epsfbox{Figures/#1.eps} }
   \caption
   {#2}
   \label{f:#1}
   \end{figure}}
\begin{document}
\title{Convergence and coupling for spin glasses and hard spheres}

\author{C\'{e}dric Chanal}

\author{Werner Krauth}

\affiliation{CNRS-Laboratoire de Physique Statistique, Ecole Normale
Sup\'{e}rieure, 24 rue Lhomond, 75231 Paris Cedex 05, France}
\date{\today}
\begin{abstract}
We discuss convergence and coupling of Markov chains, and present general
relations between the transfer matrices describing these two processes.
We then analyze a recently developed local-patch algorithm, which
computes rigorous upper bound for the coupling time of a Markov chain for
non-trivial statistical-mechanics models.  Using the \quot{coupling from
the past} protocol, this allows one to exactly sample the underlying
equilibrium distribution. For spin glasses in two and three spatial
dimensions, the local-patch algorithm works at lower temperatures than
previous exact-sampling methods.  We discuss variants of the algorithm
which might allow one to reach, in three dimensions,  the spin-glass
transition temperature. The algorithm can be adapted to hard-sphere
models.  For two-dimensional hard disks, the algorithm allows us to draw
exact samples at higher densities than previously possible.

\end{abstract}
\maketitle
\section{Introduction}

The Monte Carlo method is a fundamental computational tool in science. Its
goal is to sample configurations $x$ in a given state space from a
probability distribution $\pi(x)$. This can usually not be achieved
directly for multi-dimensional distributions.  Markov-chain Monte Carlo
methods \cite{metropolis,SMAC} overcome this problem by generating configurations $x_0,
x_1, x_2, \dots$ starting from an initial configuration $x_0$ which
belongs to a simpler distribution $\pi^0$ (often a fixed initial
condition, or some ad-hoc random choice). Configurations $x_t$ are then
generated from $x_{t-1}$ according to a stochastic algorithm which
guarantees, as time moves on, that $\pi^t$ departs from the initial
condition and converges for $t \to \infty$ towards the equilibrium
distribution $\pi^\infty \equiv \pi$. The Markov-chain approach can
be implemented for arbitrary distributions $\pi$, using for example
the Metropolis and the heat-bath algorithms.
For many applications, enormous effort has gone into designing fast
algorithms for which one reaches $\pi^t(x) \sim \pi$ for reasonable
running times $t$. In this paper, we are concerned with a related
problem: rather than to find the fastest algorithm for a given problem,
we are interested in quantifying the speed of a given Markov-chain
algorithm. This is, we want to prove after which time $t$ the sample
$x_t$ is equilibrated. It then reflects the equilibrium distribution and
no longer the initial configurations.  In many practical applications,
it is extremely difficult to decide from within the simulation whether
it has indeed equilibrated \cite{Bhatt85,Bhatt88,SMAC}. Instead, one
must validate the simulation results with other approaches, from exact
solutions to experimental data. The correct characterization of the
convergence towards equilibrium from within the simulation has remained
a serious conceptual and practical problem of the Monte Carlo method.

From a fundamental viewpoint, the problem of rigorously proving
convergence of a simulation was solved, at least in principle, through a
paradigm called \quot{exact sampling}, which allows to generate, with
Markov chains, samples $x$ directly from the equilibrium distribution
$\pi$ without any influence of the initial configuration\cite{propp}. In
practice, however, it has not been possible to implement exact sampling
for many complicated problems, as for example disordered systems,
for which standard methods of evaluating equilibration times fail.

The reason for this difficulty is as follows: Exact sampling proves
for a given Markov-chain simulation that the correlation of the initial
configuration with the configuration at time $t$ strictly vanishes. This
is done by showing explicitly that all possible initial configurations
$x_{t_0}$ yield the same output under coupled Monte Carlo dynamics. In
many simple models, one can prove this coupling property indirectly. In
general, however, one must indeed survey the entire configuration
space. This is usually too complicated to be achieved.

We recently developed a local-patch algorithm\cite{chanal} which indeed
monitors the entire configuration space of complicated systems,
even for very large sizes.  The approach uses local information,
concentrated on so-called \quot{patches}. The scale of these patches
increases during the simulation. Information on patches can then
be combined for the entire system to provide a crucial upper bound
for the (global) coupling time, and to generate an exact sample.
The algorithm was demonstrated to work for spin glasses at lower
temperatures than previous methods \cite{huber,Childs}, even though
the physically interesting regime has still not been reached yet.
The  local-patch algorithm is quite general: in addition to spin glasses,
we implement it in this paper for hard disks and improve on previous results
\cite{read_once,Kendall}. The successful application of exact sampling
to hard-sphere systems is remarquable because the configuration space
is continuous so that, naively, its complete survey appears out of reach.

\subsection{Transfer matrix}

A Markov chain is fully characterized by the so-called \quot{transfer
matrix} of transition probabilities between any two configurations $k$
and $l$. As will be illustrated shortly (\sect{coupling_one_d}) in a
specific example,  the largest eigenvalue of the transfer matrix is
$\lambda_1=1$  and the corresponding eigenvector $\Psi_1$ describes the
equilibrium state. The convergence towards equilibrium is governed by
the spectrum of the transfer matrix and by the overlap of the eigenvectors
$\Psi_k$ with the initial configuration:
\begin{equation}
\pi^t(x) = \pi(x) + \sum_{\Psi_k, \lambda_k <1} 
     \mean{\Psi_k | \pi^0 } \Psi_k(x)\ \lambda_k^t.
\elabel{exponential_convergence}
\end{equation}
In the limit of infinite simulation time, the second-largest eigenvalue
determines the exponential convergence of the probability distribution
towards equilibrium. This eigenvalue sets a time scale
\begin{equation*}
\taucorr  =1/ |\logb{\lambda_2}|, 
\end{equation*}
and the convergence is as
\begin{equation}
\pi^t - \pi \propto \expb{-t/\taucorr}\quad \text{for $t \to \infty$}.
\elabel{exponential_convergence_2}
\end{equation}
The rigorous determination of convergence properties of Markov chains
has been undertaken in many cases, from urn models to 
card-shuffling (see \cite{diaconis}), diffusion processes, and many more
(see \cite{perez}). Efficient algorithms, as for example
the bunching method \cite{SMAC} are commonly used to perform an
empirical error analysis of Monte Carlo data in more complicated cases,
where rigorous calculations are out of the question. However, these methods
are not failsafe. In practice, it is often difficult to extract $\taucorr$
from the large number of physically relevant time scales. In disordered
systems, for example, there is often no reliable way to ascertain that
the simulation has run long enough, and $\taucorr$ may be much larger
than assumed (see e.g. \cite{SMAC}, Sect. 1.5).

\subsection{Loss of correlation and exact sampling}
In the limit of infinite times $t \to \infty $, the Markov chain converges
towards the equilibrium distribution, and the positions $x_t$ become
independent of the initial condition.  The loss of correlation with
the initial condition is evident for Markov chains that couple, that
is, which for each possible initial condition $x_0$ produce the same
output $x_t$.  In many cases of interest this happens after a finite
global \quot{coupling time} $t \ge \tcoup$, which depends on the realization
of the Markov chain.  Propp and Wilson \cite{propp} realized that this
coupling property allows one to draw \quot{exact} samples from the
distribution $\pi$.

This approach, called \quot{coupling from the past}, eliminates the
problem of analyzing the convergence properties.  However, to establish
that a Markov chain has  coupled, the entire state space of the system
must be supervised. This was believed infeasible except for special problems
where the dynamics conserves a certain (partial) ordering relation on
configurations. A partial order is conserved in heat-bath dynamics of
the ferromagnetic Ising model, whereas the frustration in the spin-glass
model foils this simplification.

\section{Coupling and convergence in a one-dimensional model} 
\slabel{coupling_one_d}

We first discuss convergence and coupling for a Markov chain describing
the hopping of a single particle on a simple $N$-site lattice with periodic boundary
conditions (see \fig{one_d_single}).  In one time step, the particle hops with probability
$\tfrac{1}{3}$ from one site to its two neighbors:
\begin{equation}
   p_{k \to  k+1}  =  p_{k \to k-1} = 1/3\quad \text{(if possible)}.
   \elabel{algo_probabilities}
\end{equation}
In addition, we have  $p_{1 \to N} = p_{N \to 1}=1/3$. 
\wfigure{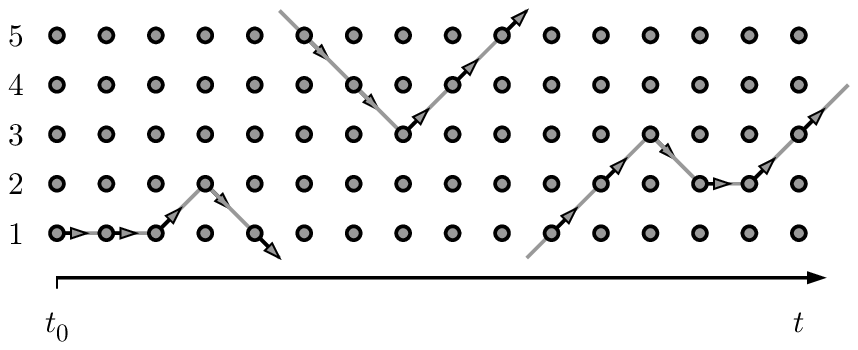}
   {A Markov chain on a five-site lattice with periodic boundary
   conditions. The particle hops from a site $k$ towards its neighbors
   with probability $1/3$ each.}

The equal hopping probabilities imply via the detailed balance condition
\begin{equation}
\pi_k p_{k \to l} = \pi_l p_{l \to k}
\elabel{detailed_balance}
\end{equation}
that the stationary probability distribution $\pi_k =1/N$ of this problem
is independent of $k$.

This system's Monte Carlo algorithm is encoded in the  $N \times N$
transfer matrix $T^{\mc}$:
\begin{equation}
T^{\mc} = 
\{p( i \to j) \} = \frac{1}{3}
\begin{pmatrix}
1      &  1       &       0 &  \cdots &       0 &       1 \\
1      &  1       &  \ddots &         &  \cdots &       0 \\
0      &  \ddots  &  \ddots &  \ddots &         &  \vdots \\
\vdots &          &  \ddots &  \ddots &  \ddots &       0 \\
0      &  \cdots  &         &  \ddots &       1 &       1 \\
1      &       0  &  \cdots &       0 &       1 &       1 \\
\end{pmatrix}.
\elabel{one_particle_MC}
\end{equation}
The eigenvalues of $T^{\mc}$ are
$\lambda^{\mc}_k=\frac{1}{3}\left(1+2\cos{\frac{2(k-1)\pi}{N}}\right),
k=1,\ldots,\text{Int}[N/2]+1$ (with multiplicities) that is, for  $N=5$, $\{1,
\frac{1 + \sqrt{5}}{6}, \frac{1 - \sqrt{5}}{6} \}$.  The largest
eigenvalue, $\lambda^{\mc}_1= 1$ corresponds to the conservation of
probabilities. By construction, it is associated with the equilibrium
solution $(\pi_{1} \TO \pi_{N})=(\frac{1}{N}\TO \frac{1}{N})$.  The second-largest
eigenvalue is $\lambda^{\mc}_2 $. For $N=5$ we have $\lambda^{\mc}_2
= \frac{1 + \sqrt{5}}{6} = 0.539$. This eigenvalue  controls the
long-time corrections to the stationary solution, which vanish as
$[\lambda^{\mc}_2]^{t} =
\expc{- t/\taucorr}$, with
\begin{equation*}
\taucorr  =1/ |\logb{\lambda^{\mc}_2}|.
\end{equation*}
We note that the time scale $\taucorr$ only describes the asymptotic
behavior of the correlation. The calculation of the time $t$ at which
the probability distribution $\pi^t$ itself is within a suitably chosen
$\epsilon$ of a the equilibrium distribution $\pi$ is more involved
(see, for example, \cite{perez,diaconis}).

\subsection{Coupling}
\slabel{coupling}
As illustrated in \fig{one_d_diffusion}, the Monte Carlo algorithm can
be formulated in terms of random maps. In our example, this means that
instead of prescribing one move per time step, as in \fig{one_d_single},
we now sample moves for all times $t$ and all sites $k$, in such
a way that the dynamics of a single particle again satisfies the
detailed balance condition of \eq{detailed_balance}.  The most natural
implementation of this approach is illustrated in \fig{one_d_diffusion}:
arrows are chosen independently for all times $t$ and all sites $k$.
At time $t_0$, for example, the particle should move down from sites $1$,
$3$, $4$ and $5$ and straight from site $2$. We can now check the outcome
of the Monte Carlo calculation.  In the example of \fig{one_d_diffusion},
from time $t_0 + 10$ on, all initial configurations of the single particle
yield the same output.  This is remarkable because,
evidently, at this time the initial conditions are completely forgotten.

The coupling time $\tcoup$ is a random variable ($\tcoup=10$ in
\fig{one_d_diffusion}) which depends on the realization of the full
Monte Carlo simulation from time $t_0$ onwards (until coupling has been
reached). The independence of random maps on different time steps implies
that the probability for not coupling vanishes at least exponentially
fast in the limit $t \to \infty$.

Under the random-map dynamics, an initial state with $N$ particles
eventually evolves into a state with one
particle (in later sections, spin-glass configurations
will take the place of the single-particle positions). More
generally, a state with $k$ configurations can evolve at each time step
into a state with $k' \le k$ configurations. \figg{one_d_diffusion}
displays a sequence of random maps and illustrates the associated
time-forward search of the coupling time.

\wfigure{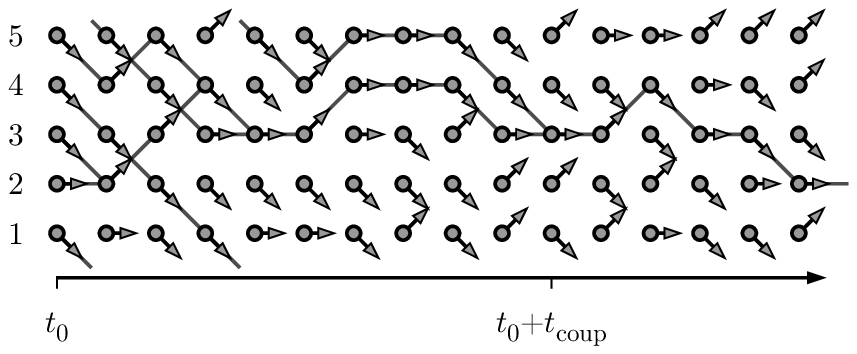} {Extended Monte Carlo simulation on $N=5$
sites. Trajectories from all possible initial configurations at $t=t_0$
are indicated.  They \quot{couple} at $t=t_0+\tcoup$. The coupling time (here
$\tcoup = 10$) depends on the realization of the Markov chain. }

This extended Monte Carlo dynamics on $k$-configuration states can again
be described by a transfer matrix:
\begin{equation}
T^\forward = 
\begin{pmatrix}
T^{1,1} & T^{2,1} & \dots  &   & \dots \\
     0  & T^{2,2} & T^{3,2}&   & \dots \\
     0  & 0       & T^{3,3}&   & \dots \\
 \dots  &         &        & \ddots & \\
     0  & 0       &        &    & T^{N,N} \\
\end{pmatrix}, 
\elabel{forward_matrix}
\end{equation}
where the block $T^{k,l}$ (of sizes $\binom{N}{k} \times \binom{N}{l}$)
concerns all the processes which lead from a state at time $t$ with
$k$ configurations to a state with $l \le k$ configurations  at
time $t+1$.  The upper left block of this matrix, $T^{1,1}$, is the
original matrix from \eq{one_particle_MC}.  As an example, we find from
\eq{algo_probabilities} the following elements of this transfer matrix:
\begin{align*} 
T^\forward\{|\circ \circ \bullet \circ \!\ \bullet \rangle &\to   |\circ \bullet \circ \bullet \!\ \circ \rangle\} = 1/9\\ 
T^\forward\{|\circ \circ \bullet \circ \!\ \bullet \rangle &\to   |\circ \circ \circ \bullet \!\ \circ \rangle\} = 1/9\\ 
T^\forward\{|\circ \circ \bullet \bullet \!\ \bullet \rangle &\to |\circ \circ \circ \bullet \!\ \circ \rangle\} = 1/27, 
\end{align*} 
etc. The matrix $T^\forward$ describes a physical system with variable
particle number (from $1$ to $N$) and a space comprising $2^N-1$ states,
the number of non-empty states in this new simulation (for a problem
of $N$ spins, the number of configurations is $2^N$ and the total
number of $k$-configuration states (states with $k$ configurations)
is $2^{2^N}-1$).

\wfigure{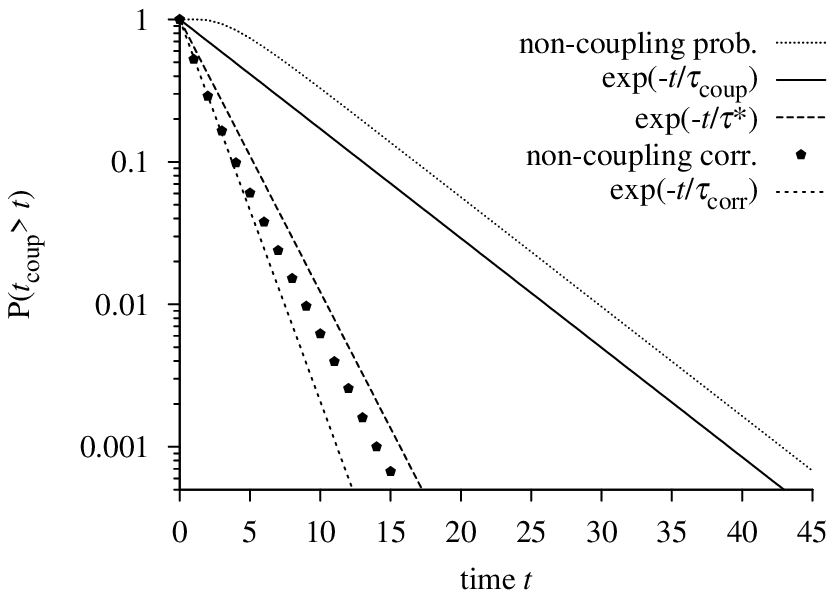}{The exact probability that the Markov chains have
not coupled by time $t$ (computed by repeated application of the forward
transfer matrix), compared to the time-scales $\taucorr$, $\taucoup$,
and $\tau^*$ (One-dimensional diffusion model with $N=5$).}

The \quot{forward} transfer matrix $T^{\forward}$  allows us to compute
the coupling probabilities as a function of time in \fig{coup_proba}.  The
matrix $T^\forward$ is block-triangular in the number of particles $(k,l)$,
with the $(1,1)$ block given by $T^{1,1}$. Therefore, all the eigenvalues
of $T^{1,1}$ are also eigenvalues of $T^\forward$.  In particular,
the largest eigenvalue of $T^\forward$ is again $\lambda^\forward_1=1$,
with corresponding right eigenvector $\tfrac{1}{5}(|\bullet \circ \circ
\circ \!\ \circ \rangle + \cdots + |\circ \circ \circ \circ \!\ \bullet
\rangle)$.  The second-largest eigenvalue of $T^\forward$ belongs
to the $(2,2)$ block and leads to the time scale of the coupling,
$\taucoup$. It is given by $\lambda^{\forward}_2 = 0.838$, larger
than $\lambda^{\text{MC}}\equiv \lambda_2^{1,1}$. This second-largest
eigenvalue $\lambda^{\forward}_2$ governs the coupling probability
$\PCAL(\tcoup)$ for large times. It follows from the block-triangular
form of the forward transfer matrix $T^\forward$ that the time scales
satisfy $\taucoup \ge \taucorr$. A  general argument allows us to
better understand this result: for any running time $t$ we may separate
all the Markov chains into those that have already coupled and those
that have not. Only the non-coupled chains (whose number vanishes as
$\expb{-t/\taucoup}$) contribute to connected correlation functions:
\begin{multline*}
\expb{-t/\taucorr} \propto \mean{\OCAL(t) \OCAL(0)}
= \sum_{\text{config. $\sigma_t, \sigma_0$}} \PCAL(\tcoup>t,\sigma_t,\sigma_0)
\OCAL(\sigma_0)\OCAL(\sigma_t) \\
+ \sum_{\text{config. $\sigma_0$}}\pi^0(\sigma_0)\OCAL(\sigma_0)\sum_{\text{config. $\sigma_t $}} \PCAL(\tcoup \le t,\sigma_t)\OCAL(\sigma_t).
\end{multline*}
Here, $\OCAL$ is an observable whose mean value is zero and $\sigma_t$
the configuration of the system at time $t$.  For the chains which have
coupled by the time $t$, $\sigma_t$ does not depend on $\sigma_0$, and
the contribution to the correlation function vanishes. For the other
chains, we find
\begin{equation}
\expb{-t/\taucorr} = 
\sum_{\text{config. $\sigma_t,\sigma_0$}} \PCAL(\tcoup>t,\sigma_t, \sigma_0)
\OCAL(\sigma_0)\OCAL(\sigma_t)
\propto \expb{-t/\taucoup}\expb{-t/\tau^*}, 
\elabel{correlation_coupling_convergence}
\end{equation}
where we suppose that even the non-coupled chains converge towards
equilibrium on a time scale $\tau^*$, and use that the probability for
a chain not to have coupled behaves as $\expb{-t/\taucoup}$ in the
long-time limit.
\eqq{correlation_coupling_convergence} shows that the difference between $\taucoup$
and $\taucorr$ is caused by the convergence taking place within non-coupling chains:
\begin{equation}
\frac{1}{\taucorr} = \frac{1}{\taucoup} + \frac{1}{\tau^*}.
\elabel{coupling_convergence}
\end{equation}
This relation is illustrated in \fig{coup_proba}  for the one-dimensional
diffusion model with $N=5$.  A later figure, \fig{spin_spin_corr},
will illustrate for the case of spin glasses the split between the
general spin--spin correlation function and that same object computed
for non-coupling chains only.

\subsection{Forward and backward coupling}
\slabel{forward_and_backward}

The probability distribution of coupling times in the forward direction
can be obtained from the transfer matrix $T^\forward$ as we discussed
in \sect{coupling}. Here we analyze the distribution of coupling times in the backward
direction for the application of the \quot{coupling from the past} protocol
which, as we will see, is the same as the one in the forward direction.
The backward coupling process leads to a generalized transfer matrix,
$T^\backward$, which again describes an extended Monte Carlo simulation.

\wfigure{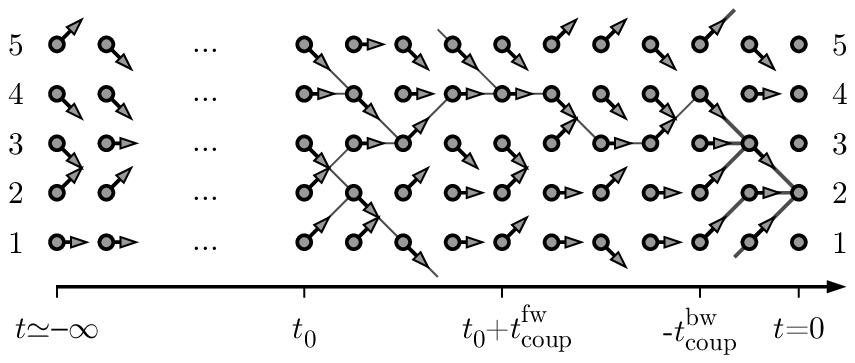} {Extended simulation on $N=5$ sites. The outcome
of this simulation,  from $t=-\infty$ up to $t=0$, is $k=2$.  It can
be obtained by backtracking from time $t=0$ to $-\tcoup^\backward$,
or by forward simulation from any $t_0 \le -\tcoup^\backward$, through
the indicated trajectories. Backtracking from sites $1,3,4$ and $5$ leads to
dead ends.}

We consider a hypothetical simulation which has run since time $t=
-\infty$ up to time $t=0$ (see \fig{one_d_cftp}).  It follows from
the discussion in \sect{coupling} that the simulation has coupled.
Furthermore, because of the infinite separation between the infinitely
remote initial condition and the final one, the
resulting configuration (at $t=0$) is in equilibrium.  But it remains
to be seen which one of the five configurations at $t=0$ is generated.
In the example of \fig{one_d_cftp}, a one-step backtrack to time $t=-1$
allows us to see that the configuration at $t=0$ can be neither $k=1$
nor $k=3$ nor $k=5$. Likewise, for $N-1$ output positions on the $N$-site ring
this back-propagation leads to a dead end, and only a single position
yields a full set of possibilities at some time $-\tcoup^{\backward}$.
Thus to find the output configuration of the simulation, one is interested
in the first time in the past for which the simulation couples, that
is one searches the \quot{backward} coupling time.  The implementation of
this backward simulation, as defined for many particles, can again
be described by a transfer matrix. For any distribution of arrows,
any occupied site $k$ at time $t$ propagates its occupation back to
all the sites at time $t-1$ which have arrows pointing towards $k$.
The matrix element of $T^{\backward}$ between two states is given by
the statistical weights of all the arrows connecting the two states.
For example, we find for $T^\backward$
\begin{align}
T^\backward\{|\circ \circ \bullet \circ \!\ \circ \rangle &\to   
        |\circ \bullet \bullet \circ \!\ \circ \rangle \} = 4/81 \notag \\
T^\backward\{|\bullet \bullet \circ \bullet \!\ \bullet \rangle &\to   
        |\bullet \circ \circ \bullet \!\ \bullet \rangle \} = 4/81 \elabel{backward_bullet_circ} \\
T^\backward\{\underbrace{|\circ \circ \bullet \circ \!\ \bullet \rangle}_{\text{time $t$}} \}&\to   
         \underbrace{|\circ \circ \circ \bullet \!\ \circ \rangle}_{\text{time $t+1$}} 
                         \} = 2/27 \notag. 
\end{align}
This is a non-trivial variant of the forward simulation as, for example,
the matrix $T^\backward$ is not block-triangular as $T^\forward$, but
it is particle--hole symmetric (as we see in the above example).

Formally, a random map $f$ (here a set of arrows) associates
configurations which are connected under the Monte Carlo dynamics.
A $k$-configuration state $|x\rangle$ is by definition a set of
configurations.  At time $t$ the state $|x\rangle$ can be associated with
the state $|y\rangle$ at time $t+1$ via the forward matrix if and only if
$|y\rangle$ is the (set) image of $|x\rangle$ by an allowed mapping $f$
(i.e. $|y\rangle=f(|x\rangle)$). The same holds for the backward matrix
but $|x\rangle$ must be the reciprocal (set) image of $|y\rangle$
(i.e. $|x\rangle=f^{-1}(|y\rangle)$).  The backward transfer matrix
$T^\backward$ manifestly differs from the forward matrix $T^\forward$.
However, we construct explicitly in \app{similarity} the similarity
transform that maps $T^\forward$ onto $T^\backward$.  This means that
\begin{equation*} P T^\forward = T^\backward P.
\end{equation*} 
(see \app{similarity}).  The similarity transformation $P$ associates a
$k$-configuration state $|x\rangle$ with the sum of states that share at
least one configuration with $|x\rangle$, included itself.  The spectrum
of the backward transfer matrix thus agrees with the one of the forward
matrix and the distribution of coupling times $\PCAL(\tcoup)$ is identical
for backward and forward dynamics.  This result is natural because
the probabilities for not coupling for  $t$ time steps are identical
in both forward and backward direction: $\PCAL(\tcoup^\backward >t) =
\PCAL(\tcoup^\forward >t)$ (see \cite{read_once}).  The probability
distribution $\PCAL(t=\tcoup^\backward)$ measures the weight of the
configuration $|\bullet \bullet \bullet \bullet \!\ \bullet \rangle $
under repeated application of the backward transfer matrix from the
configuration at $t=0$: $|\bullet \circ \circ \circ \!\ \circ \rangle +
\cdots + |\circ \circ \circ \circ \!\ \bullet \rangle$.

\subsection{Choice of random maps}
Transition probabilities of the forward transfer matrix must satisfy
the Markov chain transition probabilities for single particles, but
the choice of random maps is otherwise unrestricted.  The one-particle
sector is trivially correct for independent moves as in our diffusion
model of \sect{coupling}.  We now discuss several alternative random
maps for the one-dimensional diffusion, which may lower (or increase)
the coupling time (with, however $\taucoup\ge \taucorr$) or achieve a
rapid reduction of the number of configurations for smaller time scales.

A naive example for the one-dimensional diffusion example consists of
arrows, such as in \fig{one_d_diffusion}, but which for one time $t$
all point into the same direction, straight, up, and down, each with
probability $1/3$, so that single-particle moves satisfy the detailed
balance condition.  Evidently, this random map does not couple, and the
non-coupling Markov chains, in \eq{coupling_convergence}, converge in a
time $\tau^*=\taucorr$.  We now modify this rigid algorithm by allowing
arrows to change direction with probability $\epsilon$.  This makes the
Markov chain couple on a time scale $\sim \log{\frac{1}{\epsilon}}$,
much larger than the correlation time $\taucorr$,  for small $\epsilon$.
The choice of independent random moves ($\epsilon = 1$) is  optimal
in this class of maps, but it is not the choice minimizing $\taucoup$
among all random maps.  For example we may choose correlated moves for
selected neighboring pairs of sites say, for sites $(1,2)$ and $(3,4)$ and
let the move from site $5$ be independent (see \fig{one_d_corr_pairs}).

\wfigure{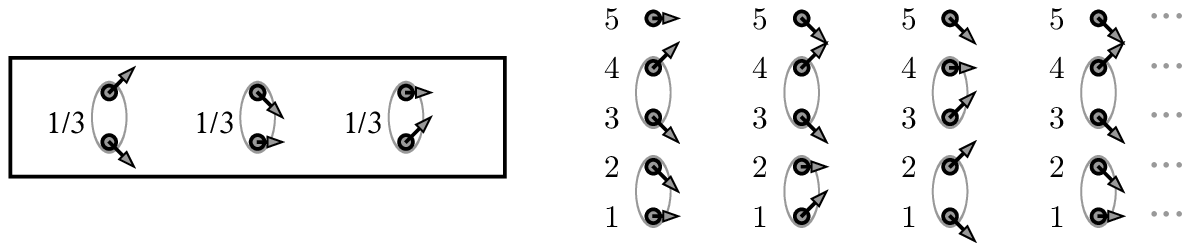} {\emph{Left:} Transition probabilities for
pairs $(1,2)$ and $(3,4)$ for the correlated random map.  \emph{Right:}
In the extended Monte Carlo simulation shown, the one-particle transition
probabilities are as in \fig{one_d_single}, but nearest-neighbor coupling
is favored.}

Elements of $T^{\map}$ are,  for example,
\begin{align*}
T^\map\{|\circ \circ \bullet \bullet \!\ \circ \rangle &\to   
         |\circ \circ \bullet \bullet \!\ \circ \rangle \} = 0  \\
T^\map\{|\circ \circ \bullet \bullet \!\ \circ \rangle &\to   
         |\circ \circ \bullet \circ \!\ \circ \rangle \} = 1/3  \\
T^\map\{|\circ \circ \bullet \bullet \!\ \circ  \rangle &\to   
         |\circ \bullet  \circ \circ \!\ \bullet \rangle\} = 1/3  \\
T^\map\{|\circ \circ \bullet \bullet \!\ \circ \rangle &\to   
         |\circ \circ \bullet \circ \!\ \bullet \rangle\} = 0.
\end{align*}
The single-particle sector of this algorithm is as before, but 
the second-largest
eigenvalue of the transfer matrix $T^\forward$ with such correlated pair
moves becomes smaller, indicating faster coupling:
\begin{equation*}
\lambda_2^{\mc}<\lambda^{\map}_2 
= 0.777 < \lambda_2^{\text{fw,indep}}=0.838.
\end{equation*}
Coupling times of both \quot{independent-arrow} random mapping and
\quot{correlated-pair} random mapping scale alike for large $N$. We note
that in applications, as in our patch algorithm of \sect{local_patch}, it
might be not so much of interest to speed up the coupling than to rapidly
decrease the number of possible configurations  at times $t < \taucoup$.
Therefore, one goal could be to decrease the eigenvalues of the matrix
$T^{k,k}, k \gg 1$, whose time scales correspond to the rapid reduction
of the number of configurations towards more manageable numbers.

\subsection{Exact sampling, coupling from the past}

As discussed in \sect{forward_and_backward}, the coupling of Markov chains
allows one to produce exact samples of the equilibrium distribution: In
the diffusion example, we were able to run the Monte Carlo simulation
backwards in time using $T^\backward$, but this matrix
can usually not be constructed.  To find the sample at time $t=0$,
one may tentatively set a time $t_0 < t$ and produce all the random maps
between time $t_0$ and $t$. One can then check explicitly whether all
the possible initial conditions at time $t_0$ have coupled, that is,
for the diffusion problem, whether the initial $N$-particle configuration
$|\bullet \bullet \bullet \bullet \!\ \bullet \rangle $ has yielded one
of the one-particle configurations.  If this goal has not been reached,
one must complement the random maps already computed with random
maps for earlier times (see \fig{one_d_cftp}). 
The one-dimensional
diffusion problem without periodic boundary conditions illustrates an
algorithm which determines the coupling time with much less effort. We
consider odd times at which only sites $1,3$ and $5$ and even time steps at which
only sites $2$ and $4$ may 
flip (see \fig{one_d_alternating}). This preserves the correct stationary
probability distribution, but the trajectories no longer cross each other
(as at time $t_0+1$ in \fig{one_d_diffusion}). As a consequence, it
suffices to follow  the two extremal configurations, which start at sites
$k=1$ and $k=N$, from time $t_0$ on in order to determine the coupling
time for a given full Monte Carlo simulation.  The multiple-particle
Monte Carlo simulation starts with the state $| \bullet \circ \circ
\circ \!\ \bullet \rangle $ until it yields a single-particle state.
The above strategy of following extremal configurations can be
applied to the ferromagnetic Ising model (but not to spin glasses,
see \cite{propp,SMAC}).  In this case, the two configurations with all
spins up and all spins down, respectively, are extremal.  This idea also
holds for the heat-bath algorithm of two-dimensional directed polymers
in a random medium.

\wfigure{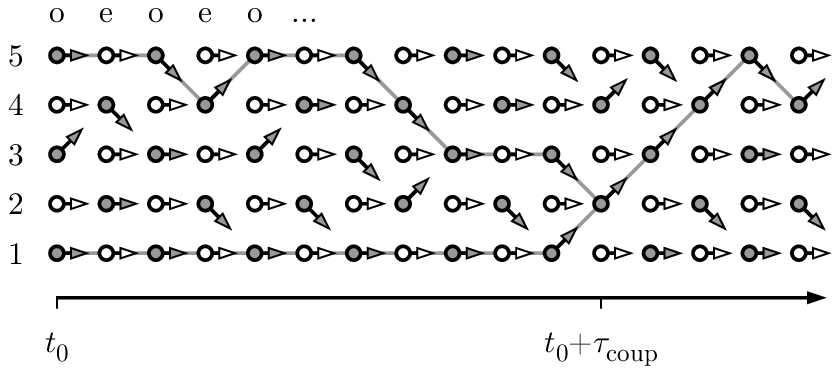}
   {Extended Monte Carlo simulation with odd (o) and even (e) time steps
   on a lattice without periodic boundary conditions. Trajectories cannot
   cross, and the coupling of the two extremal initial configurations
   (the simulations starting at time $t_0$ from sites $1$ and $5$)
   determines the coupling time.}

\section{Coupling and convergence in spin models}

We study the Edwards--Anderson $\pm J$ Ising spin glass on a
$d$-dimensional square lattice, where each site is randomly coupled to
all its $2d$ neighbors.  The energy of a configuration $\sigma=(\sigma_1,
\dots, \sigma_N)$ with $\sigma_k= \pm 1$ is:
\begin{equation*}
\HCAL(\sigma) = -\sum_{\langle i,j \rangle} J_{ij}\sigma_i \sigma_j,
\end{equation*}
where $\langle i,j \rangle$ indicates the sum over nearest neighbors.
This model has a phase transition at finite temperature in $3d$ and
at zero temperature in $2d$.  Sampling spin-glass configurations with
Markov-chain algorithms is extremely difficult in $d=3$ dimensions
below the critical temperature, but it is also non-trivial in the
two-dimensional case \cite{swendsen}. We concentrate here on the study
of a local heat-bath Monte Carlo algorithm, for which we apply the
coupling-from-the-past protocol and obtain exact samples. We note that
in two dimensions, the exact partition function of the Ising model on
a finite lattice can be determined exactly for any choice of couplings
\cite{saul,SMAC}.  This makes possible a direct-sampling algorithm,
which is completely unrelated to the material presented here, but which
we sketch, for the sake of completeness, in \app{onsager_method}.

The heat-bath Monte Carlo algorithm for spin models updates at each time
step a randomly chosen site $i$ of a spin configuration by comparing
a function of the local field on site $i$ with a uniform random number $\Upsilon_i(t)= \ran{0}{1}$:
\begin{equation}
\sigma_i(t+1) = 
\begin{cases}
1 & \text{if}\ \Upsilon_i(t) \le \glc 1 + \expa{-2 \beta h_i(t)} \grc^{-1} \\
-1 & \text{else}
\end{cases},
\elabel{heat_bath}
\end{equation}
where $h_i(t)=\sum_{j}J_{ij}\sigmavec_{j}(t)$ is the local
field on site $i$. A realization of the Markov chain
corresponds to sampling the real-valued random numbers
$\{\ldots,\Upsilon(t_0),\Upsilon(t_0+1),\ldots,\Upsilon(-1)\}$
and the random integers $\{\ldots,i(t_0),i(t_0+1),\ldots,i(-1)\}$.
The unit of \quot{physical} time (one \quot{sweep}) corresponds to $N$
individual updates. The situation is now much more complicated than for
the $1d$ diffusion, as the role of the five initial configurations
in \fig{one_d_diffusion} is taken up by the $2^N$ possible spin
configurations. To prove coupling one must  show to which configuration
they all converge at the coupling time.  The state space is huge and
one must find strategies to avoid enumerating and surveying $2^N$
configurations.

\subsection{Partial-survey approximation}
\slabel{partial_survey}

In \cite{chanal}, we presented an exact-sampling algorithm which
works down to quite low temperatures in the two-dimensional Ising spin
glass, and which is also operational in three dimensions.  We found
that practically the same results could be obtained by starting the
simulation at time $t_0$ not from all the $2^N$ initial configurations,
but from a more manageable number $\NCAL(t_0)$ of randomly chosen
configurations.  We show in \fig{ultra_naive_16x16_0.5} that such
\quot{partial survey} calculations yield useful lower bounds for the
coupling time scale $\taucoup$.  Each curve in the figure represents
the mean number of distinct configurations remaining after coupled Monte
Carlo simulations (that is, with the same random numbers  $(\Upsilon, i)$
for all configurations) for different values of $\NCAL(t_0)$. Increasing
$\NCAL(t_0)$ within this partial-survey approximation naturally improves
the lower bound on the coupling time but, in practice, the value obtained
saturates quite quickly.
\wfigure{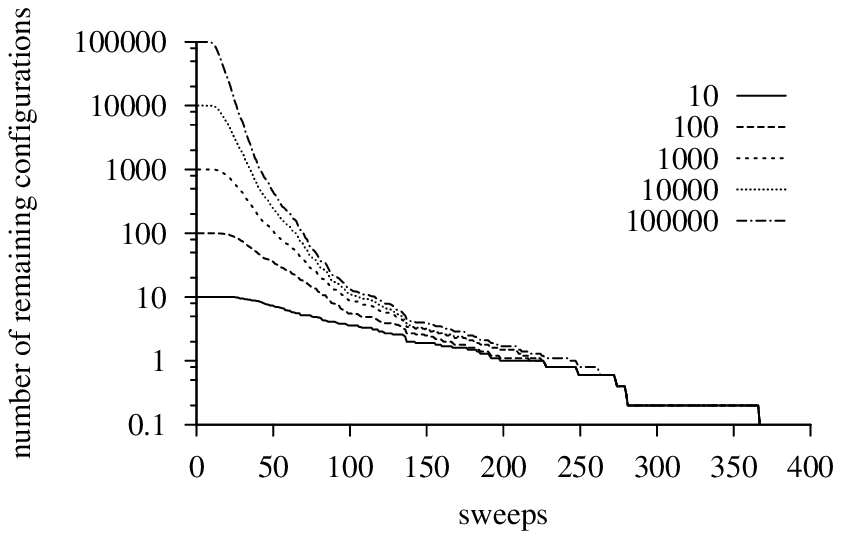}
   {Number of configurations of the partial survey approximation with
   $\NCAL(t_0)$ random initial configurations for the Ising spin glass on a
   $16\times16$ lattice at temperature $\beta=0.5$.  We average over $10$
   choices of $J_{ij}$, and use the same values of $J_{ij}$ and the same
   random numbers $\Upsilon_i, i$ for all initial configurations.}

\subsection{Correlation functions}
As discussed previously,  $ \taucoup $ is always larger than $\taucorr$
because only non-coupling chains contribute to correlation functions
(see \eq{coupling_convergence}).
\wfigure{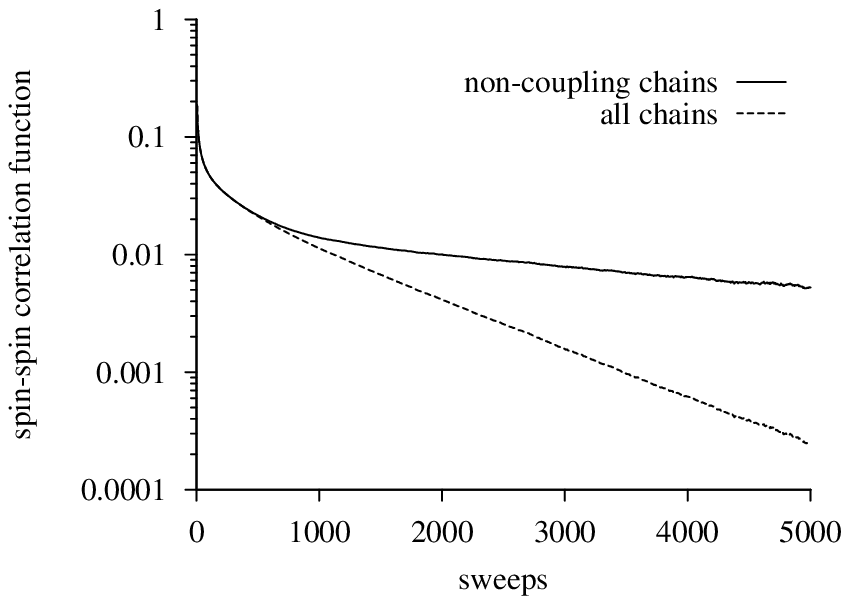}
   {Spin--spin correlation function for all Markov chains and for the
   non-coupling Markov chains only ($\beta=1$, random initial conditions)
   (Two-dimensional $8\times8$ spin glass at $\beta=1$).}
To again illustrate the relation between coupling and convergence
times, we separate in \fig{spin_spin_corr} non-coupling chains from
the calculation of the spin--spin correlation function of a $8\times8$
spin glass at inverse temperature $\beta=1$.  Indeed, even if the chain
has not coupled, the configurations $\sigma_t$ may lose the dependence
on the initial configuration $\sigma_0$.

\section{Coupling for hard-sphere systems}
In this section we discuss the application of the \quot{Coupling from
the past} protocol to hard-sphere systems.  The study of Monte Carlo
algorithms for hard-sphere systems goes back a long time, as the
Metropolis algorithm was first implemented for hard disks, that is,
two-dimensional spheres\cite{metropolis}.  Even today, the physics of
the hard-disk system is not well understood, and Monte Carlo algorithms
have not been developed as successfully as, say, for the Ising model.
In this very constrained system, the estimation of correlation times is
quite controversial, especially at high densities\cite{bernard_2009},
and rigorous results from exact-sampling approaches would be extremely
welcome.

We first discuss the birth--death formulation of the Markov-chain
Monte Carlo algorithm for this system and then compute lower bounds
on coupling times using the partial-survey algorithm. Its empirical
coupling time saturates (for increasing $\NCAL(t_0)$) to much smaller
values than the coupling times obtained by the summary-state method
\cite{read_once,Kendall}.  This suggests that these previous algorithms
are not optimal, an impression which is confirmed by our local-patch
algorithm of \sect{patch_disk}.

The partition function of hard spheres in the grand-canonical ensemble,
with fugacity $\lambda$, is given by a weighted sum over legal
configurations of spheres:
\begin{equation}
\mathcal{Z} =\sum_{N=0}^{+\infty} \int\!\! 
    d^{2N} \!\sigma^{(N)}~ \lambda^{N} ~\  \Theta \glb \sigma^{(N)}\grb.
\elabel{partition_function}
\end{equation}
Here, configurations of $N$ spheres are written as: 
\begin{equation*}
\sigma^{(N)} =
\{(x_1,y_1),(x_2,y_2),\ldots,(x_N,y_N)\},
\end{equation*}
where $(x_k,y_k)$ denotes the centers of the spheres. In
\eq{partition_function}, $\Theta(\sigma^{(N)})$ equals one if spheres
of the configuration $\sigma^{(N)}$ do no overlap and zero otherwise.
We again use periodic boundary conditions.\\

\subsection{Birth--death algorithm for hard spheres}

The spatial Poisson birth--death process allows us to apply coupling
from the past to hard-sphere systems (see \cite{Kendall, read_once}): 
Disks of radius $r$ arrive (\quot{are born}) randomly
on a two-dimensional unit square with constant rate $\lambda$. Once born,
they disappear (\quot{die}) with unit rate.

The probability for a disk to arrive within an infinitesimal time $dt$
in a small box of area $dS$ centered at $(x,y)$ is $\lambda dtdS$.
This disk is added to the configuration only if it overlaps  with no other
disk present. Each disk disappears with probability $dt$ within the time
interval $dt$.  Sphere are added at point $(x,y)$ or removed from the
configuration $\sigma^{(N)}$ according to the detailed balance condition.
With the notation $\sigma^{(N+1)}
= \sigma^{(N)} \cup \{(x,y)\}$ we have:
\begin{align*}
\PCAL(\sigma^{(N)} \rightarrow \sigma^{(N+1)}) \pi(\sigma^{(N)}) 
&= \lambda ~ \Theta(\sigma^{(N+1)}) \pi(\sigma^{(N)}) \\
&= \pi(\sigma^{(N+1)}),  \\
\PCAL(\sigma^{(N+1)} \rightarrow \sigma^{(N)}) 
    \pi(\sigma^{(N+1)}) &= 1 \times \pi(\sigma^{(N+1)}).
\end{align*}

In \fig{birth_death_w}, we illustrate the time-evolution of accepted and
rejected birth-death events on a one-dimensional hard-sphere problem,
starting from an empty initial condition at time $t_0$.

\wfigure{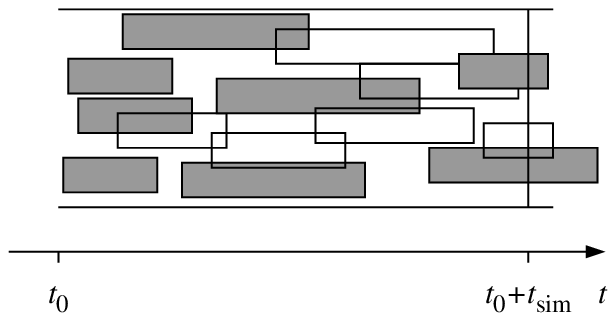}{Simulation of the birth--death algorithm for
one-dimensional \quot{spheres} in a box. The simulation starts at time $t_0$
and stops at $t_0 + t_{\text{sim}}$ with $N=2$ spheres. Transparent spheres
are rejected because they overlap with spheres already present.}

In the hard-sphere algorithm, the probability distribution of time intervals
between successive births is an exponential with parameter $\lambda$:
$\PCAL(\tau_b) = \lambda e^{-\lambda \tau_b}$.  In \fig{birth_death_w},
the life time of a sphere is represented by a horizontal extension of
the box, irrespective of whether it has been accepted or not (the
vertical dimension denotes the diameter). Life times are exponentially
distributed as well. For the exponential distribution, the time before
the next death of a system of $N$ spheres follows an exponential
distribution with parameter $N$.  Likewise the time before any event,
(birth or death), follows an exponential distribution with parameter
$\lambda + N$.  The probability for the next event to  be a birth is
then $\frac{\lambda}{\lambda + N}$.

\subsection{Coupling and partial survey approximation}
\slabel{hard_disk_coup}

Coupling from the past applies to hard-sphere systems even though
the space of configurations is continuous (unlike in lattice
simulations). To apply the protocol, one considers a time evolution,
as in \fig{birth_death_w}, but stretching back to time $t=-\infty$. Two
special aspects must now be handled:

First, we must determine which boxes (corresponding to spheres) are indeed
placed (\quot{True}), and which ones are rejected (\quot{False}). This
is difficult to decide at high density. However, in the low-density
case presented in \fig{time_space}, several spheres are \quot{True},
simply because they do not overlap with already present \quot{True}
or \quot{False} spheres.  This allows the status of other spheres to
be fixed and, finally, the configuration to be constructed.  In the
limit $N \to \infty$, the approach works up to a constant density
\cite{luby}. This density is much higher than the density $\propto
1/N$ direct-sampling algorithm can achieve\cite{SMAC}. This approach
\cite{Kendall,read_once} is equivalent to deciding whether a given spin
is up or down in the \quot{summary state} algorithm for Ising systems
\cite{huber,Childs}, which, in the thermodynamic limit works down to a
fixed constant temperature.

Second, one must fix the initial condition at time $-T$, because spheres
born at times smaller than $-T$ may still be alive at time $-T$.  This is
solved through the sampling of a second time, $T_{\text{start}}$, after
which we know that all spheres present at time $-T$ have disappeared. The
time interval $\Tstart + T$  is sampled as the maximum of $N_{\max}$
life times, where $N_{\max}$ is an upper bound on the number of spheres
in a legal configuration.

\wfigure{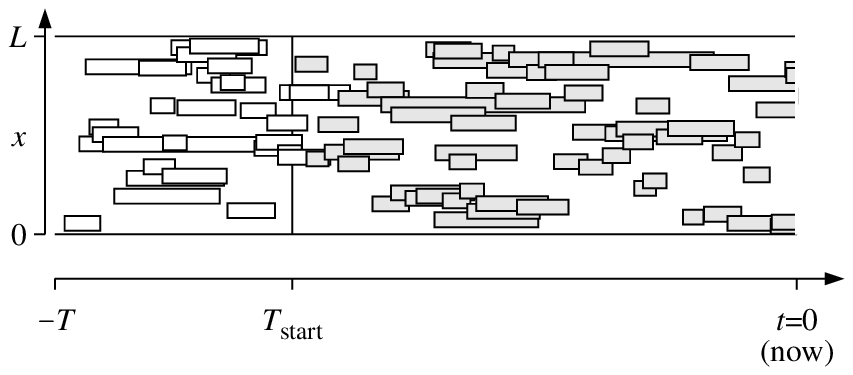}{Time evolution of a one-dimensional birth--death
simulation in a box of size $L$. All spheres correspond to rectangles
whose horizontal extension indicates their life  time.  From any possible
cut of \quot{True} boxes at $\Tstart$ ($7$ boxes actually cut the line,
so there are $\le 2^7$ possible cuts) one can deduce the output at time
$t=0$, as in \fig{birth_death_w}.}

\figg{time_space} sketches the time evolution of a Monte Carlo simulation
for the one-dimensional hard-sphere problem which has started at time
$t=-\infty$. Boxes are drawn starting from time $-T$, but the simulation
is picked up at time $\Tstart$.  It is straightforward to complement
the simulation shown (between times $-2T$ and $-T$, for example), in
case it does not couple in the interval shown. However, we must show
that it couples between $-2T$ and $-T$ or at least results in less than
$N_{\max}$ spheres. In the Monte Carlo simulation in \fig{time_space},
the status of the boxes at later times can be easily decided, because
at later times all spheres belong to clusters which are disjoint from
the initial condition.  However, this possibility disappears at higher
densities. A simple example of this is shown in \fig{example_BD}. As
one cannot decide on the status of the initial sphere (which crosses
the line at $\Tstart$), we should initialize the simulation with the
two configurations, one corresponding to a \quot{True} state and one
to a \quot{False} state.  After several steps of the time evolution,
we arrive in both cases at the same physical configuration (the two dark
spheres, which are both \quot{True}).

\wfigure{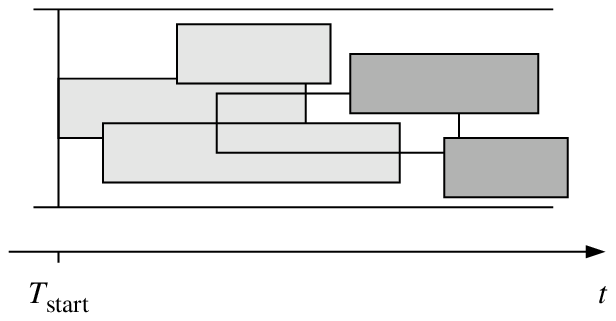}
   {Example of a time evolution of the one-dimensional birth--death
   simulation which where no single sphere can be decided
   independently. Starting with all possible choices of the initial
   configuration at $t=\Tstart$ allows to prove coupling. (The two
   dark spheres are \quot{True}, while the transparent sphere must be
   \quot{False}.)}

For all times $t> \Tstart$, we consider the set $\Tries(t)$ of all
\quot{True} or \quot{False} spheres crossing the time line at $t$ (see
\fig{time_space}).  From the set $\Tries$, one can in principle construct
all the possible initial configurations, but their number remains huge. As
in the spin-glass case, we may also select $\NCAL(\Tstart)$ among these
configurations, and propagate these. This is again the partial survey
approximation.  In \fig{patches_vs_Wilson} we compare average lower
bounds on the coupling time from this approximation with results from
the summary state algorithm\cite{read_once}.  In the time-evolution of
\fig{time_space}, one can determine the number of remaining configurations
at any time $ t>\Tstart $ and detect when exactly the coupling occurs.

\wfigure{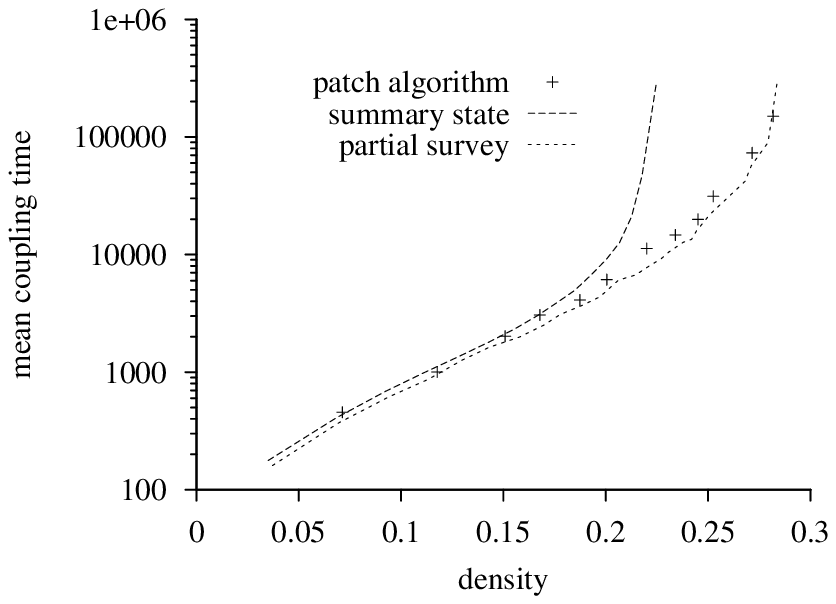}{Coupling times of the summary-state algorithm
for hard disks \cite{Kendall,read_once} compared to the local-patch
algorithm.  Lower bounds are provided by the partial-survey approximation.
The disks' radius is $r=0.04$ in a unit square box with periodic boundary
conditions, so that there are $\sim 60$ disks at density $\eta=0.3$.}

\section{Local-patch algorithm}
\slabel{local_patch}
In the present section, we discuss our local-patch algorithm, which
performs the heat-bath dynamics for a general $d$-dimensional Ising
spin glass on an $N$-site hyper-cubic lattice. This algorithm allows
us to control all the $2^N$ initial configurations even for very
large lattices and to eventually prove that the system has coupled.
The Python script implementing this algorithm has less than 300 lines. It
is available electronically and a listing of the code is contained in
\app{listing_python}.

\subsection{Patches}
The (non-rigorous) partial-survey algorithm of \sect{partial_survey}
determines the coupling time for a subset of all the configurations
at time $t$. The (rigorous) patch algorithm, in contrast, works
with a superset of all configurations at time $t$: by restricting the configurations to
the smaller region of a patch, one severely limits their number, at the price of
introducing compatibility problems between neighboring patches. For
the two-dimensional spin glass, we use $N$ rectangular patches of same
shape and orientation, with $M$ sites, and initially at $t=t_0$, we have
$2^M$ spin configurations  on each patch.  Likewise,  the set of
global spin configurations is broken up into a list $[S_1(t) \TO S_N(t)]$
of sets $S_k(t)$ of spin configurations restricted to patches $k$. We
can recover a superset $\Omega(t)$ of all relevant spin configurations from the direct product
\begin{equation} \Omega(t) =  S_1(t) \otimes S_2(t) \otimes \cdots
\otimes S_N(t)/\text{(compat)}.  
\elabel{Omega}
\end{equation}
Here, each configuration of $\Omega(t)$ is pieced together from
configurations on all patches, with compatible spins on all lattice sites.
(Two compatible spin configurations, on patches $k$ and $l$, are shown
in \fig{patches_schema}). On large lattices, the direct product in
\eq{Omega} can be performed only if the number of spin configurations
per patch is small.  If there is only one configuration per patch,
we can construct a unique global configuration on the whole lattice.

For each time step $t$ of the heat-bath algorithm, we choose a random
lattice site $i$ and a random number $\Upsilon = \ran{0}{1}$ and
then update the spin $\sigma_i$ for all configurations on all patches
containing $i$. The site $i$ may be in the center of a patch $k$ (all the
neighbors of site $i$ also belong to $k$, as in \fig{patches_schema}). In
this case, each configuration of $S_k(t)$ yields one configuration
of $S_k(t+1)$. Several configurations in $S_k(t)$ may yield the same
configuration in $S_k(t+1)$, so that the size of $S_k$ does not increase
in this case.  If the site $i$ is on the boundary of a patch $l$,
we only know upper and lower bounds for the field on the site $i$ and,
depending on the value of the random number $\Upsilon$, may be unable to
update $\sigma_i$.  In this case, we add two configurations to the set
$S_k(t+1)$, corresponding to $\sigma_i=-1$ as well as $\sigma_i=+1$. The
set $S_k(t+1)$ may then contain more configurations than $S_k(t)$.

\wfigure{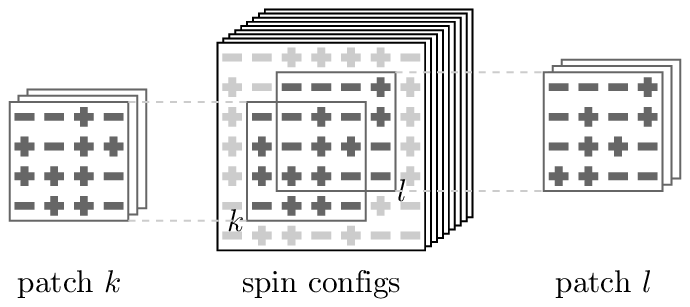}{Two patches, $k$ and $l$, with a pair of
compatible spin configurations.}

\subsection{Compatibilities, pruning}
\slabel{pruning}

Besides updating configurations on patches, we also perform
a \quot{pruning} operation: \figg{patches_schema} presents two
\quot{compatible} configurations on patches $k$ and $l$. These could
possibly be pieced together into a global configuration, together with
configurations on other patches. On the other hand, if the set $S_l$
contains no configuration compatible with a configuration $\sigmavec$ on
patch $k$, we can eliminate (prune) $\sigmavec$ from $S_k$.  Pruning may
be implemented through a dictionary (hash table), using as \quot{key}
the part of the patch configuration in the overlap region between $k$
and $l$, and as \quot{value} the list of patch configurations sharing
this key (see \fig{dictionary}). This is programmed very easily in the
Python programming language (see \app{listing_python}).

\wfigure{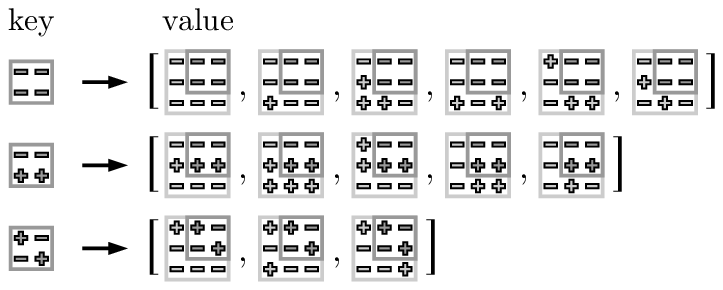} {A dictionary (hash table) associating keys
(configurations in the overlap region between patches $k$ and $l$) to
values (lists of patch configurations with the given key in patch $k$).}

Pruning can be iterated until all the sets $S_k(t)$ are pairwise
compatible. To achieve this goal, it suffices to prune nearest-neighbor
patches only. We have found it useful to perform one pruning operation
for each pair of nearest-neighbor patches after a certain number of
updates (see \fig{patches_straight_16x16_1.0}).  The average number
of configurations per patch saturates to a value which depends on
the temperature and also the size of the patch. This is due to the
balance between the decrease of the number of configurations induced
by the coupling and its increase caused by the noise at the patch
boundaries. This noise is reduced through the crucial pruning step of
the algorithm.

\wfigure{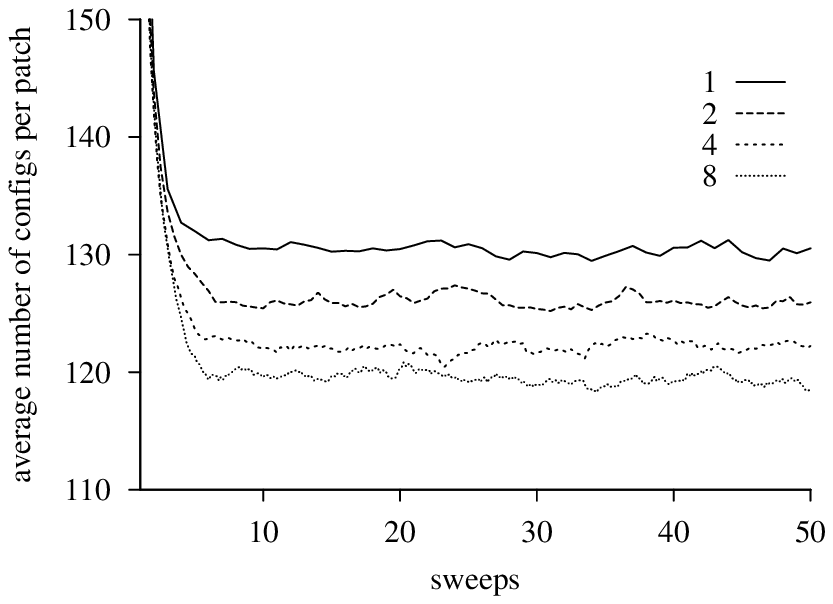} 
{Average number of configurations per patch \emph{vs.} time, at inverse
temperature $\beta=1$ in the two-dimensional Ising spin glass on a
$16\times 16$ lattice at temperature $\beta=1$. For a given patch size
(here $3\times 3$) the number of configurations per patch saturates to
a value that  depends on the number of prunings per sweep (here: from  one
to eight). Each pruning is done once for all nearest-neighbor patches.}

\subsection{Merging of patches}

As illustrated in \fig{patches_straight_16x16_1.0}, the number of
configurations per patch does not necessarily drop to one at large
times, even if the underlying heat-bath dynamics couples. The entropy
per spin is smaller for larger patches, because the influence of the
boundaries is reduced. However, one cannot start the computation with
large patches because of the large number of possible configurations.
A merging procedure allows to increase the patch size in
a rigorous way.  Merging is implemented analogously to pruning: for
overlapping patches $k$ and $l$, dictionaries are again computed with
the same keys and values (see again \fig{patches_schema}). For a given
key configuration on the overlap region, we assemble the corresponding
values in the dictionary of patch $k$ with all corresponding values in
the dictionary of patch $l$. All these couples of configuration must be
taken into account for the larger patch $k\cup l$. The merging of the
configurations on neighboring patches can be implemented very efficiently
in the Python programming language (see \app{listing_python}).  In our
computations, we start with small square patches, say of size $3\times
3$, and then pass to the size $4\times3$, after a few sweeps, then to
size $4\times 4$, etc.  An analogous procedure is followed in higher
dimensions. Results obtained with this \quot{jump-start} approach are
shown in \fig{fig_jump_32x32_0.5_5gen} for the two-dimensional Ising
spin glass at temperature $\beta=0.5$, with a disorder average performed
over about $100$ samples.  
\wfigure{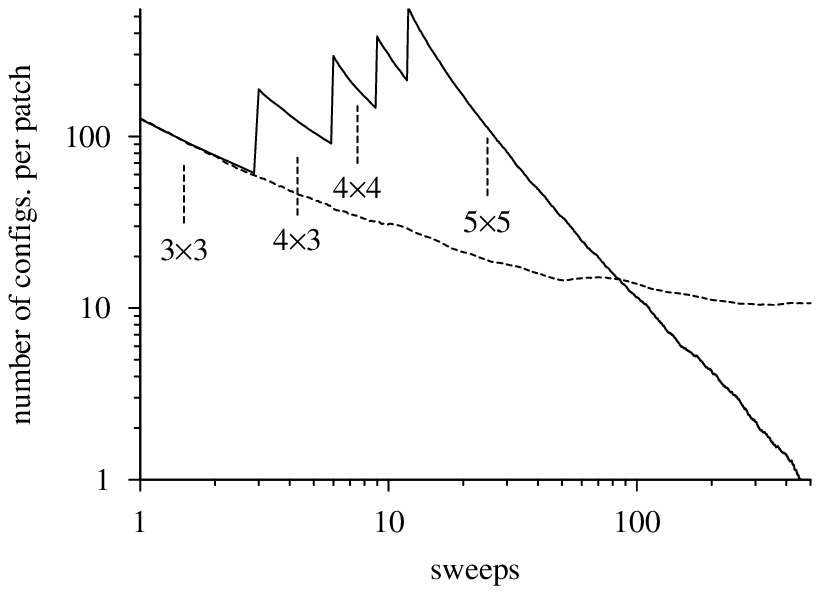} {Number of
configurations per patch for the $32\times 32$ spin glass at temperature
$\beta=0.5$. A simulation with constant $3\times 3$ patches is compared
to the result of a \quot{jump-start} procedure $3\times3 \to 4\times 3
\to 4\times4 \to \dots \to  5\times 5$. All results are averaged over
$100$ samples.}

\subsection{Memory of compatibilities, variants}
In the patch algorithm, the pruning procedure detects inconsistent
configurations in a particular patch. Two configurations on different
patches are considered compatible if their spins match in the overlap
region (at time $t$). More generally, we can keep track of the past
evolution of patch configurations and may then declare them compatible
only if they have matched for all times up to $t$. Otherwise, they cannot
belong to a unique global spin configuration.

In \cite{chanal}, the bipartite nature of the square lattice was used
to update one entire sub-lattice at a time. In this approach, only
one sub-lattice is stored at a time. This allows one to start with
larger patches, but the compatibilities between configurations are
less well conserved. By contrast, in the present algorithm, we keep the
information on both sub-lattices, and one of the sub-lattices is the past
configuration. Two configurations are compatible in this new version
if they are compatible on both sub-lattices.  Likewise one can use past
values of a configuration $\sigma(t)$ to restrict its compatibilities
with configurations on other patches.

Other generalizations are more straightforward, one can for example
optimize the shape of patches in order to minimize the number of spins
on the boundary, and work with more than $N$ patches in order to increase
the chance for detecting incompatibilities of configurations.

\subsection{Exact sampling for hard spheres with local patches}
\slabel{patch_disk}

\wfigure{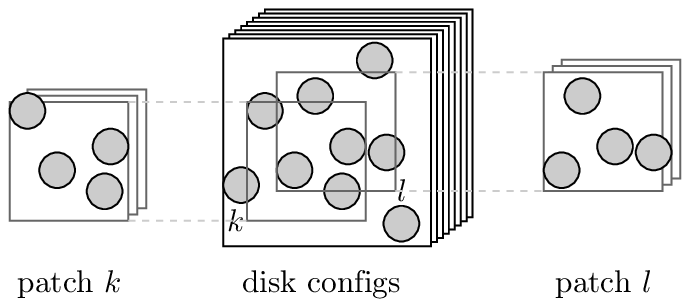}{Breaking up disk configurations into patches,
with two patches $k$ and $l$ shown.}

In this section, we adapt the local-patch algorithm for the classical
model of hard disks in a two-dimensional box with periodic boundary
conditions.  As mentioned, it works for large system sizes at higher
density than previous method of exact sampling.

In \sect{hard_disk_coup}, we introduced the set $\Tries(t)$ of  all disks
the dynamics has tried to add in the box and which have not disappeared
at time $t$. The coupled Monte Carlo simulations start with all possible
configurations that are allowed by the set $\Tries$ at time $\Tstart$.
Because of possible overlap between disks some of the $2^{\#\Tries}$
possible configurations are invalid but there may be far too many of
them in practice.  To reduce the number of configurations which must be
handled, we introduce a regular square lattice with $N$ sites covering the
simulation box.  Likewise, a superset of all feasible configurations on a
patch at time $\Tstart$ can be deduced from the set $\Tries$, restricted
to disks (True or False) with centers inside the patch. From then on,
whenever disks appear in the simulation box, we can decide whether they
are accepted on a particular patch configuration by checking overlaps
into the patch only. Disks that disappear are simply removed from all the
concerned patch configurations. At birth time, if the disk to be placed
on a patch configuration may overlap with a disk outside the patch, the
configuration is split into two: one configuration with the new disk
and one without (as for spin systems). To detect and prune irrelevant
configurations we check that the updated sets of configurations are
compatible with other patches by a pruning procedure analogous to the
one of \sect{pruning}.  After several updates, the pruning is performed
for most of overlapping patches several times.

\wfigure{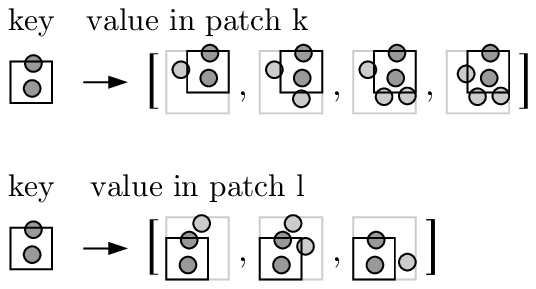}{The pruning of a pair of patches via
the construction of dictionaries.  Each dictionary associates keys
(configurations in the overlap region) to values (lists of patch
configurations with the given key) (compare with \fig{dictionary}).
The merging of patches $k$ and $l$ would lead in this example to
$4\times3=12$ configurations on the combined patch.}

For any pair of overlapping patches $k$ and $l$, the pruning eliminates
patch configurations with are inconsistent with all other configurations
on a neighboring patch.  As in the  spin glass case, this process can be
implemented with dictionaries (hash tables) (see \fig{dictionary_disks})
and can be used to merge configurations on neighboring patches into
larger local configurations.

\figg{patches_vs_Wilson} displays the mean coupling times of a hard-disk
birth--death simulation for several choices of the fugacity $\lambda$.
The radius of the disks is $ r=0.04$ in a unit square box with periodic
boundary conditions.  Results are compared to the partial-survey
approximation algorithm, with $\NCAL(\Tstart)=1000$ initial configurations
and to the results of the summary-state algorithm.

We concentrated on determining the coupling times of the birth--death
dynamics for hard disks. However, the regime of operation of this
algorithm is far in the liquid phase (see \fig{patches_vs_Wilson}), and
the physically interesting regime, around the liquid--solid transition
density, $\eta \simeq 0.71$ \cite{SMAC} is still out of reach for exact
sampling methods. For hard disks, it remains a challenge to set up a
working partial-survey algorithm with correlation times comparable to
those of the usual Metropolis algorithm \cite{bernard_2009}.

\section{Conclusion}

In this paper, we have discussed exact-sampling algorithms which
allow one to totally eliminate the influence of the initial condition
from a Markov-chain Monte Carlo simulation. This overcomes one of the
main limitations of the method, namely the rigorous estimation of the
correlation time.  We discussed central subjects, such as the relation
between coupling times and convergence times, in a simple example of
one-dimensional diffusion, before applying them to Ising spin glasses
and to hard-sphere simulations.  Algorithmically, the exact-sampling
framework obliges one to follow the entire state space of a system. In
the absence of simplifications, such as the half-order discussed
in \sect{coupling_one_d}, this can be done approximately through a
partial survey of $\NCAL(t_0)$ initial conditions. One can also restrict
the configuration in size onto so-called patches, thereby restricting
their number.  A superset of the set of global configurations can in
principle  be reconstituted from the patches.  This is easier when the
patches are large, and we showed how  pruning and merging operations
allow one to increase the size of patches during the simulation and to
finally prove coupling.  Our exact-sampling  algorithm  works both for
spin glasses and hard-disk systems, and were able to go to lower
temperatures, and higher densities than previous methods.

The partial survey algorithm, which can be implemented easily, allowed
us to prove that our local-patch algorithm is optimal for the local
dynamics for both spin-glass and hard-disk systems. We have provided a
number of new idea in order to allow exact-sampling methods to reach
the phases transitions of the three-dimensional Ising spin glass and
the critical density of hard disks.

\appendix
\section{Similarity of forward and backward transfer matrices}
\slabel{similarity}
Forward and backward transfer matrices completely describe the coupling
dynamic of general Markov chains (on a finite state space), that is
their elements are the coupled transition probabilities between  sets of
configurations.  Forward and backward matrices represent \quot{extended}
Monte Carlo dynamics, in the two time directions.  These two formulations
are equivalent, even though the forward dynamics, starting from $t=0$,
does not generate exact samples. Here,  we demonstrate similarity
between forward and backward transfer matrices, and construct the
similarity transformation between the two.

Let $\Omega$ be the finite space of configurations of the problem
of interest.  $\Omega$ may contain all $N$ positions on the $N$-site
diffusion problem or the $2^N$ configurations of a $N$-site spin system.

The $k$-configuration states build up an \quot{extended} state space. They
provide a natural basis for the forward and backward matrices.  This basis
is $2^\Omega-\emptyset$, the set of non trivial parts of $\Omega$.
For any state $I\in 2^\Omega-\emptyset$ we define $\tilde{I}$ as the set
of states of the basis $J$ that has at least one configuration in common
with $I$ ($J \in \tilde{I} \Leftrightarrow I\cap J \ne \emptyset$).
The similarity matrix $P$  is then defined as:

\begin{equation}
P|I\rangle := \sum_{J \in \tilde{I}} |J\rangle
\elabel{similarity}
\end{equation}

A random mapping---arrows for the case of 1$D$-diffusion---is a mapping on
$\Omega$: $f:\Omega \rightarrow \Omega$.  It defines a time step of the
Markov chain for every configuration and satisfies $\PCAL(f(i)=j) =
p(i\rightarrow j)$ and its weight is noted $w(f)$.  In the case of
1$d$-diffusion with independent arrows,  or any \quot{independent}
random map in general, we naturally define the weight of the random
map as a product of elements of the Monte Carlo transfer matrix as
\begin{equation*}
w(f) = \Pi_{i\in\Omega} p(i\rightarrow f(i)).
\end{equation*}

The forward matrix associates any state $I$ to all states that it is 
connected to by a mapping:
\begin{equation*}
T^\forward|I\rangle = \sum_{f\ \text{rand. map}} w(f) |f(I)\rangle.
\end{equation*}
Using \eq{similarity} we find
\begin{equation}
P\ T^\forward|I\rangle = \sum_{f\ \text{rand. map}} w(f) \sum_{J \in \tilde{f(I)}}|J\rangle.
\elabel{PF}
\end{equation}

The backward matrix $T^\backward$ has different rules but we will
show that the similarity $P\ T^\forward =  T^\backward
P$ holds.  Using a random map $f$, a state $J$ at time $t$ evolves to
another state $K$ at time $t+1$ in the backward process if and only if
$f^{-1}(K)=J$. For example in the 1D-diffusion a hole goes to a hole
and a particle goes to a particle.  Therefore:
\begin{equation*}
T^\backward |J\rangle = \sum_{f\ \text{rand. map}} w(f) \sum_{K, f^{-1}(K)=J} |K\rangle
\end{equation*}
and finally:
\begin{equation}
T^\backward \ P|I\rangle = \sum_{f\ \text{rand. map}} w(f)\sum_{J \in
\tilde{I}} \sum_{K, f^{-1}(K)=J} |K\rangle = \sum_{f\ \text{rand. map}}
w(f) \sum_{K, f^{-1}(K)\in \tilde{I}} |K\rangle.
\elabel{BP}
\end{equation}
In fact \eqtwo{BP}{PF} are equivalent because $f^{-1}(K)$ overlaps $I$
if and only if $K$ overlaps $f(I)$ ($f^{-1}(K)\cap I \neq \emptyset
\Leftrightarrow K\cap f(I) \neq \emptyset $).  This proves the similarity
of the backward and forward matrices.

\section{Exact sampling of two-dimensional spin glass using analytic
solution.}
\slabel{onsager_method}

In this appendix, we sketch for completeness an unrelated direct-sampling
algorithm for the two-dimensional Ising spin glass. To generate exact
samples, this algorithm does not use Markov chains. It rather relies on
the fact that the partition function of the two-dimensional Ising model
or of one sample of the spin glass on a planar lattice with $N$ sites
can be expressed as the square root of the determinant of one $4N\times
4N$ matrix (for open boundary conditions) or of four such matrices
(for periodic boundary conditions) \cite{saul}. This relation has been
much used in the recent literature, in order to study the physics of the
two-dimensional Ising spin glass at low temperature \cite{loebl,marinari}.
The partition function yields the thermodynamics of the system, but the
knowledge of entire configurations gives for example access to complicated
spatial configuration functions.

The sampling algorithm for two-dimensional spin-glass configurations
constructs the sample one site after another.  Let us suppose that
the gray spins in the left panel of \fig{onsager_direct} are already
fixed, as shown. We can now set a fictitious coupling $J_{ll}^*$ either
to $-\infty$ or two $+ \infty$ and recalculate the partition function
$Z_\pm$ with both choices. The statistical weight of all configurations in the
original partition function with spin \quot{$+$} is then given by
\begin{equation}
\pi_+ = \frac{Z_+ \expb{\beta J_{kl}}}
{Z_+ \expb{\beta J_{kl}} + Z_- \expb{-\beta J_{kl} }}.
\elabel{direct_sampling}
\end{equation}
and this two-valued distribution can be sampled with one random
number.  \eqq{direct_sampling} resembles the heat-bath algorithm of
\eq{heat_bath}, but it is not part of a Markov chain:  After
obtaining the value of the spin on site $k$, we keep the fictitious
coupling, and add more sites.  Going over all sites, we can generate
direct spin-glass samples at any temperature. We note that this algorithm
is polynomial, and the effort is basically  temperature-independent,
both for the two-dimensional Ising model and the Ising spin glass (see
also \cite{Krauth09}).

\wfigure{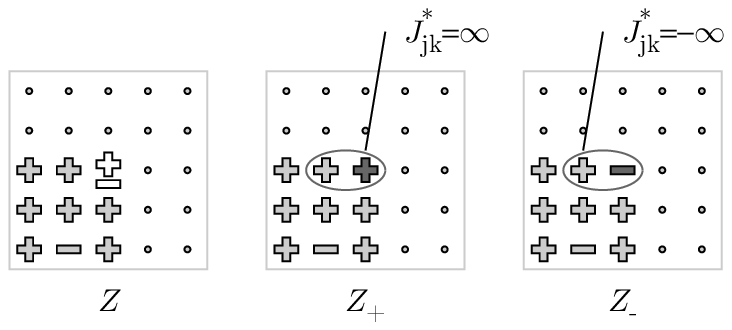}{One iteration in the direct-sampling algorithm
for the two-dimensional Ising model. The probabilities $\pi(\sigma_k=+1)$
and $\pi(\sigma_k=-1)$ (with $k$ the central spin) are obtained from the
exact solution of the Ising model with fictitious couplings $J_{jk}^*=
\pm \infty$.}

\section{Listing of Python code}
\slabel{listing_python}
The following Python code has produced all the spin-glass data presented
in this article.  An analogous program was used for the hard-disk system.
An electronic version of the code is available from the authors.
\begin{lstlisting}[caption=pruning-ND.py]
#!/usr/bin/python
##---------------------------------------------------------------------
## PROGRAM : pruning_ND.py
## PURPOSE : This program performs the heat-bath and the pruning on a
##           N-dimensional hypercubic lattice with periodic boundary
##           conditions and with cuboid patches.
##           Version with jump-start capability.
## OUTPUT  : mean number of configurations per patch vs. time 
##           (can be modified)
## VERSION : 04-OCT-2009
## AUTHOR  : W. Krauth, C. Chanal
## LANGUAGE: Python 2.5
##---------------------------------------------------------------------
from random import uniform, randint, seed, shuffle, choice
from operator import itemgetter
import time, math, os, sys
##---------------------------------------------------------------------
## Sample geometry
##---------------------------------------------------------------------
def torus_neighbors(N_dim,L):
   N = L**N_dim
   site_dic = {}
   coord_dic = {}
   for j in range(N):
      j_d = j
      x_d = j//(L**(N_dim-1))
      coord = [x_d]
      for d in range(N_dim-1): # this loop does not run anything for N_dim=1
         j_d = (j_d-x_d*L**(N_dim-d-1))
         x_d = j_d//(L**(N_dim-d-2))
         coord.append(x_d)
      coord.reverse() # optional, to set the usual order of the directions
      coord=tuple(coord)
      site_dic[coord] = j
      coord_dic[j] = coord
   nbr = []
   for j in range(N):
      coord = list(coord_dic[j])
      nbr_list = []
      for d in range(N_dim):
         coord[d] = (coord[d]+1)%L
         coord_p = tuple(coord)
         nbr_list.append(site_dic[coord_p])
         coord[d] = (coord[d]-1+L)%L    
      for d in range(N_dim):
         coord[d] = (coord[d]-1+L)%L
         coord_p = tuple(coord)
         nbr_list.append(site_dic[coord_p])
         coord[d] = (coord[d]+1)%L         
      nbr_list = tuple(nbr_list)
      nbr.append(nbr_list)
   nbr = tuple(nbr)
   return nbr,site_dic,coord_dic
##---------------------------------------------------------------------
## Initial patch geometry
##---------------------------------------------------------------------
def patch_set(N_dim,L,M,site_dic,coord_dic):
   patch = []
   for j in range(N):
      coord = list(coord_dic[j])
      dummy_list = []
      for k in range(M**N_dim):
         coord_p = []
         for d in range(N_dim):
            l = (k//(M**d))%M
            coord_p.append((coord[d]+ l)%L)
         coord_p = tuple(coord_p)
         dummy_list.append(site_dic[coord_p])
      patch.append(tuple(dummy_list))
   patch = tuple(patch)
   return patch
##---------------------------------------------------------------------
## Neighbor relations on patches
##---------------------------------------------------------------------
def nbr_patch_set(example_patch,nbr):
   nbr_patch = []
   for j in example_patch:
      dummy = []
      for k in nbr[j]:
         if k in example_patch: dummy.append(example_patch.index(k))
      dummy = tuple(dummy)
      nbr_patch.append(dummy)
   nbr_patch = tuple(nbr_patch)
   return nbr_patch
##---------------------------------------------------------------------
## Initial patch configurations
##---------------------------------------------------------------------
def patch_init(patch):
   def bin(n,conf_length):
##
## convert n to binary number with conf_length digits
##
      q = -1
      bin_conf = ''
      n_digits = 0
      while q != 0:
         q = n//2
         r = n%2
         bin_conf = `r`+bin_conf
         n = q
         n_digits = n_digits+1
      n_digits = conf_length-n_digits
      for i in range(n_digits):
         bin_conf = `0`+bin_conf
      return bin_conf
   configs = []
   dummy_list = []
   number = len(patch[0])
   for j in range(2**number):
      x = bin(j,number)
      dummy_list.append(x)
   for i in range(len(patch)):
      configs.append(set(dummy_list))
   return configs
##---------------------------------------------------------------------
## Updating configurations on all patches
##---------------------------------------------------------------------
def confs_update(N_dim,configs,nbr_patch,Jij,patch,i_site,Upsilon):
   for k in range(N):
      if i_site in patch[k]:
         dummy = list(patch[k])
         pos=dummy.index(i_site)
         llist = [patch[k][x] for x in nbr_patch[pos]]
         Jij_list = [Jij[(i_site,m)] for m in llist]
         config_k = configs[k] 
         config_kp = set([])
         for c in config_k:
            field_char = itemgetter(*nbr_patch[pos])(c)
            field_sum = sum((2*eval(m)-1)*J for (m,J) in zip(field_char,Jij_list))
            field_min = field_sum-2*N_dim+len(field_char)
            field_max = field_sum+2*N_dim-len(field_char)
            b_one = c[:pos]+'1'+c[pos+1:] # configuration with '1' at position 'pos'
            b_zero = c[:pos]+'0'+c[pos+1:] # configuration with '0' at position 'pos'
            if Upsilon < phplus[field_min]: 
               config_kp.add(b_one)
            elif Upsilon > phplus[field_max]: 
               config_kp.add(b_zero)
            else:
               config_kp.add(b_zero)
               config_kp.add(b_one)
         configs[k] = config_kp
   return configs
##---------------------------------------------------------------------
## Pruning of two patches
##---------------------------------------------------------------------
def prune_pair(config_k,config_l,KEY_k,KEY_l):
   f = itemgetter(*KEY_k)
   Dic_k = {}
   for x in config_k:
      a = f(x)
      if Dic_k.has_key(a): Dic_k[a].add(x)
      else: Dic_k[a] = set([x])
   set_k = set(Dic_k.keys())
   f = itemgetter(*KEY_l)    
   Dic_l = {}
   for x in config_l:
      a = f(x)
      if Dic_l.has_key(a): Dic_l[a].add(x)
      else: Dic_l[a] = set([x])
   set_l = set(Dic_l.keys())
   set_kl = set.intersection(set_k,set_l) 
   config_k = set()
   config_l = set()
   for x in set_kl:
      config_k.update(Dic_k[x])
      config_l.update(Dic_l[x])
   return config_k,config_l
##---------------------------------------------------------------------
## Merging of two patches
##---------------------------------------------------------------------
def merge_pair(config_k,config_l,KEY_k,KEY_l,KEY_add):
   f = itemgetter(*KEY_k)
   Dic_k = {}
   for x in config_k:
      a = f(x)
      if Dic_k.has_key(a): Dic_k[a].add(x)
      else: Dic_k[a] = set([x])
   set_k = set(Dic_k.keys())
   f = itemgetter(*KEY_l)    
   Dic_l = {}
   for x in config_l:
      a = f(x)
      if Dic_l.has_key(a): Dic_l[a].add(x)
      else: Dic_l[a] = set([x])
   set_l = set(Dic_l.keys())
   set_kl = set.intersection(set_k,set_l) 
   config_kl = set() #merged set
   for x in set_kl:
      for y in Dic_k[x]:
         for z in Dic_l[x]:
            zprime = ''.join(itemgetter(*KEY_add)(z))
            config_kl.add(y+zprime)
   return config_kl
##---------------------------------------------------------------------
## main program starts here
##---------------------------------------------------------------------
#seed(13)
beta =0.25 # inverse of temperature
##---------------------------------------------------------------------
## heat bath definitions (see SMAC Fig. 5.20, and SMAC eq. 5.18)
##---------------------------------------------------------------------
N_dim = 3
phplus = {}
for d in range(N_dim+1):
   field = 2*d
   phplus[field] = 1/(1+math.exp(-2*field*beta))
   phplus[-field] = 1/(1+math.exp(2*field*beta))
##---------------------------------------------------------------------
## loop over samples starts here
##---------------------------------------------------------------------
L = 6
os.system('echo `hostname` `date`')
print  beta, L, ' beta L'
for nsamp in range(100):
   N = L**N_dim
   M = 2 # M^N_dim is the initial size of patches
   over_min = (M-1)*M**(N_dim-1) # minimum overlap for the pruning 
   del_t = 40 # time lap on each patch size
   n_gen = 7 # number of generations 
   t_max = 2000 # total number of sweeps
   N_frac = 6 # do N/N_frac spin updates between prunings
   nbr,site_dic,coord_dic = torus_neighbors(N_dim,L)
##---------------------------------------------------------------------
## Jij: a dictionary (i,j) -> J_ij
##---------------------------------------------------------------------
   Jij = {}
   for k in range(N):
      for d in range(N_dim):
         Jij[(k,nbr[k][d])] = choice([-1,1])
         Jij[(nbr[k][d],k)] = Jij[(k,nbr[k][d])]     
   patch = patch_set(N_dim,L,M,site_dic,coord_dic)
   nbr_patch = nbr_patch_set(list(patch[0]),nbr)
   permut = [k for k in range(N)]
   configs = patch_init(patch)
##---------------------------------------------------------------------
## Loop over generations in the jump-start procedure
##---------------------------------------------------------------------
   for iter1 in range(n_gen):
      i_dir = iter1 % N_dim
      tot_it = 0.
      patch_size = len(patch[0])
      print patch_size, ' size of patch'
      while (tot_it < del_t and iter1 < n_gen-1) or (iter1 == n_gen-1 and \ 
                   tot_it +del_t*iter1< t_max): 
         quality = sum([len(configs[k]) for k in range(N)])/float(N)
         sys.stdout.flush()
         print tot_it+del_t*iter1, quality
         #if (iter1==n_gen-1): N_frac=18 # do N/N_frac spin updates between prunings
         for iter3 in range(N/N_frac):
            tot_it += 1./N
            i_site = randint(0,N-1)
            Upsilon = uniform(0,1)
            configs = confs_update(N_dim,configs,nbr_patch,Jij,patch,i_site,Upsilon)
##---------------------------------------------------------------------
## Pruning starts here
##---------------------------------------------------------------------
         shuffle(permut)
         for kk in range(N):
            k = permut[kk]
            for ll in range(kk+1,N):
               l = permut[ll]
               inter_set = set(patch[k]) & set(patch[l])
               if len(inter_set) >= over_min:  # minimum overlap
                  KEY_k = [list(patch[k]).index(i) for i in inter_set]
                  KEY_l = [list(patch[l]).index(i) for i in inter_set]
                  configs[k],configs[l] = prune_pair(configs[k],configs[l],KEY_k,KEY_l)
         if (quality==1): break
      if (quality==1):break
##---------------------------------------------------------------------
## Merging starts here: producing new patches, and computing configurations on them
##---------------------------------------------------------------------
      patch_new = []
      configs_new = []
      for k in range(N):
         KEY_add = []
         dummy_new = list(patch[k])
         l = nbr[k][i_dir] # l is the neighbor of k in the 'i_dir' direction
         for x in patch[l]:
            if x not in patch[k]:
               dummy_new.append(x)
               KEY_add.append(list(patch[l]).index(x))
         patch_new.append(tuple(dummy_new))
         inter_set = set(patch[k]) & set(patch[l])
         KEY_k = [list(patch[k]).index(i) for i in inter_set]
         KEY_l = [list(patch[l]).index(i) for i in inter_set]
         config_k_merge = merge_pair(configs[k],configs[l],KEY_k,KEY_l,KEY_add)
         configs_new.append(config_k_merge)
##---------------------------------------------------------------------
## Merging (final steps): computing neighbor relations on the new patches 
##---------------------------------------------------------------------
      patch = tuple(patch_new)
      nbr_patch = nbr_patch_set(list(patch[0]),nbr)
      configs = configs_new
      if i_dir == N_dim-1 :
         M += 1
         over_min = (M-1)*M**(N_dim-1)


\end{lstlisting}

\end{document}